# Machine learning based white matter models with permeability: An experimental study in cuprizone treated in-vivo mouse model of axonal demyelination

**Ioana Hill**[1,a], **Marco Palombo**[1,a], **Mathieu Santin**[b,c], **Francesca Branzoli**[b,c], **Anne-Charlotte Philippe**[b], **Demian Wassermann**[d,e], **Marie-Stephane Aigrot**[b], **Bruno Stankoff**[b,f], **Anne Baron-Van Evercooren**[b], **Mehdi Felfi**[g], **Dominique Langui**[g], **Hui Zhang**[a], **Stephane Lehericy**[b,c], **Alexandra Petiet**[b,c], **Daniel C. Alexander**[a], **Olga Ciccarelli**[h], **Ivana Drobnjak**[a]

**Affiliations:**

[a] Centre for Medical Image Computing and Dept of Computer Science, University College London, London, UK

[b] Institut du Cerveau et de la Moelle épinière, ICM, Sorbonne Université, Inserm 1127, CNRS UMR 7225, F-75013, Paris, France

[c] Institut du Cerveau et de la Moelle épinière, ICM, Centre de NeuroImagerie de Recherche, CENIR, Paris, France

[d] Université Côte d'Azur, Inria, Sophia-Antipolis, France

[e] Parietal, CEA, Inria, Saclay, Île-de-France

[f] AP-HP, Hôpital Saint-Antoine, Paris, France

[g] Institut du Cerveau et de la Moelle épinière, ICM, Hôpital Pitié Salpêtrière, 47 Boulevard de l'Hôpital, 75013 Paris

[h] Dept. of Neuroinflammation, University College London Queen Square Institute of Neurology, University College London, London, UK

## Abstract

The intra-axonal water exchange time $\tau_i$, a parameter associated with axonal permeability, could be an important biomarker for understanding and treating demyelinating pathologies such as Multiple Sclerosis. Diffusion-Weighted MRI is sensitive to changes in permeability, however, the parameter has so far remained elusive due to the intractability of the mathematical models that incorporate it. Machine learning based computational models can potentially be used to estimate such parameters, and recently, for the first time, a theoretical framework using a random forest (RF) regressor suggests that this is a promising new approach for permeability estimation. In this study, we adopt such an RF approach and experimentally investigate its suitability as a biomarker for demyelinating pathologies through direct comparison with histology. For this, we use an in-vivo controlled cuprizone (CPZ) mouse model of demyelination with available ex-vivo electron microscopy (EM) data.

Using simulations closely matching our in-vivo data, we show that our imaging protocol has good sensitivity to $\tau_i \leq 400$ms. We optimise the performance of our RF model by choosing the most informative b shells with respect to $\tau_i$ and show that it is necessary to include a combination of short and long diffusion times to maximise sensitivity to the exchange time. We establish a benchmark performance for our model by testing it on noise-free simulations, where we find strong correlations between the predicted and ground truth parameters (*intra-axonal volume fraction f*: $R^2=0.99$, $\tau_i$: $R^2=0.84$, *intrinsic diffusivity d*: $R^2=0.99$). We find that for realistic levels of noise such as SNR=40, as present in our in-vivo data, the simulation performance is affected, however, the parameters are still well estimated (f: $R^2=0.99$, $\tau_i$: $R^2=0.68$, d: $R^2=0.99$). We apply our RF model on in-vivo data consisting of a cohort of 8 CPZ mice and 8 healthy age-matched wild-type (WT) mice. We find that the model estimates sensible microstructure parameters matching values found in literature. We validate these using histology and find a strong correlation between the in-vivo RF estimates of $\tau_i$ and the EM measurements of myelin thickness ($\varrho_{\tau i} = 0.82$), and between RF estimates and EM measurements of intra-axonal volume fraction ($\varrho_f = 0.98$). Furthermore, we show using NODDI ODI and EM that our estimates are not confounded with the effects of dispersion and axonal swelling. When comparing the exchange time in CPZ and WT mice we find a statistically significant decrease in $\tau_i$ in all three regions of the corpus callosum (splenium/genu/body) of the CPZ group ($\mu_{\tau i}$=310ms/330ms/350ms) compared to the WT group ($\mu_{\tau i}$=370ms/370ms/380ms), in line with our expectations that $\tau_i$ is lower in regions where the myelin sheath is damaged, as axonal membranes become more permeable. Overall, these results demonstrate, in simulations and in vivo, the suitability of machine learning based compartment models with permeability as a potential biomarker for demyelinating pathologies such as Multiple Sclerosis.

---

[1] These authors contributed equally to this work.

# 1. Introduction

The intra-axonal water exchange time $\tau_i$, a parameter associated with axonal permeability, is an important microstructural property of the tissue, which has been linked with the condition of the myelin sheath surrounding the axons in brain tissue (Nilsson et al., 2013b). Several neurological conditions such as Multiple Sclerosis (MS) cause a breakdown of the myelin sheath through a process known as demyelination, which may lead to an increase in the exchange time as the intra-axonal water molecules encounter less barriers. Changes in permeability have also been linked with pathologies such as Parkinson's disease (Volles et al., 2001) or cancer (Hu et al., 2006), leading to a widespread interest in developing permeability-based biomarkers. Due to its sensitivity to the motion of water molecules within tissue, Diffusion-Weighted MRI (DW-MRI) is potentially able to estimate $\tau_i$. However, measuring it has been problematic due to the intractability of the mathematical expressions which accurately incorporate this parameter into analytical models.

So far, the analytical models that incorporate permeability rely on assumptions that are either too simplistic (Callaghan, 1997, Codd and Callaghan, 1999, Vangelderen et al., 1994) or do not hold in human tissue (Grebenkov et al., 2014, Kärger et al., 1988). The Kärger model (Kärger et al., 1988) is the most widely used analytical model that incorporates permeability (Nilsson et al., 2010, Stanisz et al., 2005, Lätt et al., 2009). However, its assumptions do not hold in white matter and the model was shown to fail when applied to highly permeable tissue (Fieremans et al., 2010). An alternative analytical model is the apparent exchange rate (AXR) imaging, however, AXR cannot differentiate between changes in permeability and intra-axonal volume fraction, and requires a specialised imaging protocol (Lasič et al., 2011, Nilsson et al., 2013a).

Computational models bypass the need for analytical models and incorporate permeability by creating a mapping between simulations of the DW-MRI signal and the ground truth microstructure parameters. Nilsson et al. (2010) use Monte Carlo simulations with known ground truth parameters including permeability to generate a synthetic library of DW-MRI signals. Given a previously unseen signal, they estimate permeability via a nearest-neighbour algorithm. However, their approach requires new libraries to be generated for each acquisition protocol and fibre orientation, and the nearest-neighbour algorithm in general does not have a good generalisation capacity.

Recently, Nedjati et al. (2017) introduce for the first time a machine learning approach using a random forest and a database of rotationally invariant features derived from the raw DW-MRI signals. Their model uses a random forest for better generalisation and is independent of the fibre orientation. The novel random forest model is shown to outperform the Kärger analytical model on synthetic and in-vivo human data (Nedjati et al., 2017). However, their in-vivo approach is tested only qualitatively on just two MS patients. Furthermore, Nedjati et al. (2017) hypothesise that $\tau_i$ is linked with demyelination in MS lesions, however, they do not show whether other underlying processes such as axonal swelling or orientation dispersion affect the estimates. Here, we aim to address these issues.

The aim of this study is to experimentally test a machine learning based computational model with permeability using a highly controlled cuprizone-treated, in-vivo mouse model of demyelination, with a direct comparison to histology. We adopt the random forest framework introduced in Nedjati et al (2017) to estimate tissue microstructure parameters. Prior to our in-vivo experiments, we use simulations representative for our mouse data to investigate the sensitivity of the PGSE protocol used to acquire the in-vivo data to $\tau_i$, and select the most informative b shells (i.e. b values and directions) with respect to this parameter. We additionally establish a benchmark performance for our model by testing its performance on simulations. To test the in-vivo performance of the model, we use a cohort of 16 mice (8 cuprizone, 8 wild-type) with DW-MRI scans and histology data. Our demyelination model allows us to investigate the direct correlation between the estimated exchange time and histological measurements of myelin thickness. Furthermore, we investigate the potentially confounding effects of dispersion and axonal swelling to eliminate any potential bias in our predictions of the exchange time. Finally, we analyse the correlations between the predictions of our model and histology data.

## 2. Methods

### 2.1 Diffusion imaging protocol

We use the same DW-PGSE protocol for the synthetic and in-vivo data, optimised to maximise signal reconstruction accuracy under realistic time constraints (Filipiak et al., 2019). Our imaging protocol has 25 shells, each with one b=0 measurement and a different combination of diffusion gradient strength G and diffusion gradient separation Δ as summarised in *Table 1*. The resulting protocol has 345 measurements in total, diffusion gradient duration δ=5ms, $|G_{max}|$=500mTm$^{-1}$ and shell b-values as shown in *Table 1*. Additional protocol details are as follows: TE=33.6ms, TR=2s, FOV=16x16mm, matrix size = 160x160, number of slices=5, slice thickness=0.5mm.

| Δ(ms)<br>G (mT/m) | 10.8 | 13.1 | 15.4 | 17.7 | 20 | #grad dirs |
|---|---|---|---|---|---|---|
| **150** | 358 | 445 | 533 | 620 | 707 | **16** |
| **200** | 620 | 775 | 930 | 1086 | 1241 | **16** |
| **300** | 1384 | 1733 | 2083 | 2432 | 2781 | **8** |
| **400** | 2489 | 3110 | 3731 | 4352 | 4973 | **11** |
| **500** | 3892 | 4862 | 5833 | 6803 | 7773 | **13** |

**Table 1.** *DW-PGSE parameters with the corresponding nominal b-values in s/mm$^2$.*

### 2.2 Synthetic data

To construct our computational model, we first generate two synthetic databases: one formed of simulated DW-MRI signals and one formed of rotationally invariant features derived from the simulated signals. Each entry in the database corresponds to a unique digital phantom which mimics the in-vivo data and for which the ground truth microstructure parameters are known. Each synthetic database is used to train a machine learning algorithm, here a random forest, to build a mapping between the signal or features and the corresponding ground truth microstructure parameters.

#### 2.2.1. Synthetic signals database

We use Monte Carlo simulations of the DW-MRI signal to build our synthetic training database. The signals are generated using the open source Camino simulation framework (Cook et al., 2006; http://camino.cs.ucl.ac.uk) together with the imaging protocol in *Table 1*. Each simulated signal corresponds to a digital phantom which mimics the in-vivo mouse brain data introduced in *Section 2.3*. The digital phantoms are represented by synthetic substrates that model white matter as a collection of 100,000 non-abutting, parallel cylinders with gamma-distributed radii, a common choice in the brain literature (Aboitiz et al., 1992). The cylinders are randomly packed in the substrates as described in Hall and Alexander (2009), with example substrates shown in *Figure 1*. We construct a database of 11,000 unique tissue substrates and their corresponding DW-MRI signals by randomly sampling from a range of histologically plausible substrate parameters for white matter tissue (Aboitiz et al., 1992, Barazany et al., 2009). A white matter synthetic substrate is defined through five parameters: the mean $\mu_R \in [0.2,1]$μm and the standard deviation $\sigma_R \in [\min(0.1, \mu_R/5), \mu_R/2]$μm of the axon radii distribution, the intra-axonal volume fraction $f \in [0.4, 0.7]$, the intra-axonal exchange time $\tau_i \in [2, 1000]$ms and the intrinsic diffusivity $d \in [0.8, 2.2]\mu m^2 ms^{-1}$. To ensure the convergence and the high precision of the simulated signals, we generate our synthetic database using 100,000 spins and 2,000 time steps (Hall and Alexander, 2009). The permeability of a substrate is specified within the Camino simulation framework via the probability parameter $p$. This parameter expresses the probability that a spin steps through a membrane encountered during the random walk (instead of always being reflected backwards as it is the case for impermeable substrates). The probability $p$ is related to the permeability $k$ through the expression:

$$p = k\sqrt{6 \frac{\delta t}{d}},$$

where $d$ is the intrinsic diffusivity and $\delta t$ is the temporal resolution. This expression is obtained by combining the Monte Carlo step length equation $s = \sqrt{6d\delta t}$ (Hall and Alexander, 2009) with the transition probability equation as derived by Regan et al (Regan and Kuchel, 2000). Here, we measure permeability $k$ via the intra-axonal water exchange time $\tau_i$, which is inversely related to $k$ through the expression $k = \frac{R}{2\tau_i}$, where R is the axon radius (Fieremans et al., 2010).

To maximise the performance of our machine learning regressor, we aim to build a training database that resembles as closely as possible the in-vivo data. For this, we generate an additional set of synthetic signals to account for the noise present in the in-vivo data. We add Rician noise with a standard deviation σ corresponding to an SNR of 40, which reflects the noise level of the b=0 images with the longest **Δ.**

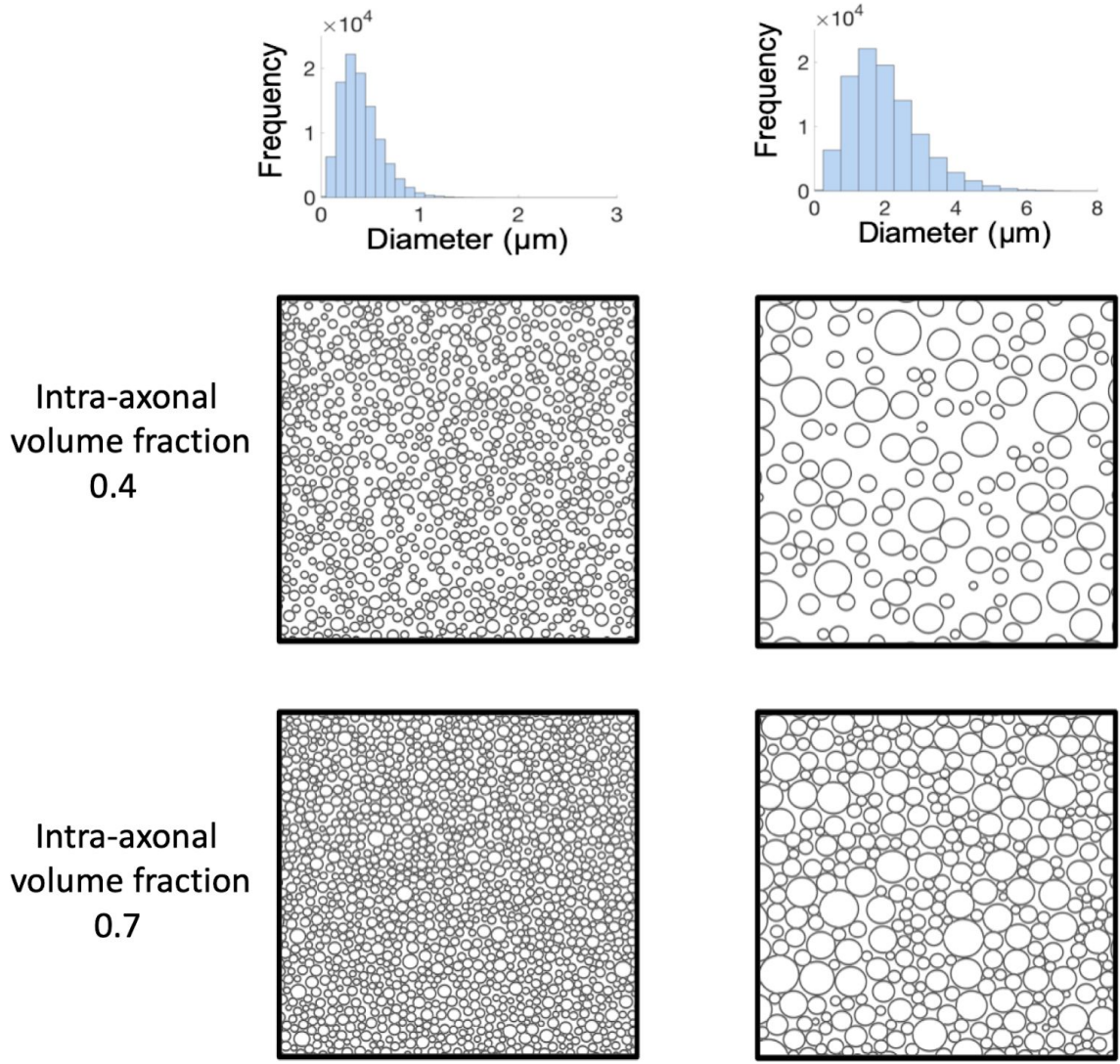

*Figure 1:* *Histograms of two example axon diameter distributions used to generate synthetic substrates for our Monte Carlo simulations (first row). Figure also shows four digital tissue substrates corresponding to the two example axon diameter distributions and two different intra-axonal volume fractions: 0.4 (second row) and 0.7 (third row).*

*2.2.2 Synthetic features database*

In addition to the signals database, we construct a rotationally invariant feature database as suggested in Nedjati et al. (2017). We compute the diffusion tensor (DT) and the 4$^{th}$ order spherical harmonic (SH) fit for each b shell from the synthetic signals using the Camino toolkit (Cook et al., 2006). We then derive 15 rotationally invariant features for each b shell and build an equivalent rotationally invariant synthetic database. The first five signal-derived features are calculated from the DT fit and are the three eigenvalues $\lambda_1, \lambda_2, \lambda_3$, the mean diffusivity (MD) and the fractional anisotropy (FA). The remaining ten features are derived from the SH fit: the mean, peak, anisotropy, skewness and kurtosis of the apparent diffusion coefficient together with the peak dispersion and simple combinations of the first, second and fourth order spherical harmonics (Nedjati et al., 2017).

## *2.3 Mouse data*

### *2.3.1 In-vivo data acquisition*

We imaged sixteen 8-week old C57BL/6J female mice. All animal experiments were performed in accordance with the European Council Directive (88/609/EEC). Eight mice were fed 0.2% cuprizone for 6 weeks, which corresponds to a demyelination without recovery phase, and eight healthy age-matched wild-type (WT) mice of the same background were fed a normal chow diet and used as controls. All mice are scanned on a BrukerBioSpec 11.7T scanner (for the open access database see (Wassermann et al., 2017)) using the protocol described in *Section 2.1*.

We post-processed the images by correcting for eddy currents using FSL-eddy (Smith et al., 2004). No motion artifacts were observed. We restrict our analysis to white matter voxels within the corpus callosum (CC) as they match the assumption of non-abutting parallel cylinders in our learnt white matter model. To select the CC voxels, we compute maps of linearity ($C_L$), planarity ($C_P$) and sphericity ($C_S$) (Westin et al., 2002) from the DT fit to the lowest b-value shell with SNR 40. We created the CC maps by selecting the voxels with $C_L$>0.3, $C_P$<0.4, $C_S$<0.5 and fractional anisotropy (FA)>0.3.

### *2.3.2 Histology samples*

The WT (n=8) and CPZ (n=8) animals are sacrificed by deep anaesthesia and perfused intracardially with 1% paraformaldehyde and 2.5% glutaraldehyde in phosphate buffer 0.12 M, pH 7.4 at the end of the 6-week CPZ treatment. The extracted brains are then post-fixed overnight at 4°C in the same fixative and rinsed in phosphate buffer. Ten 100μm-thick sagittal sections are cut with a vibratome (Thermo Scientific Microm HM 650 V Vibration microtome). The very first section closest to the brain midline is considered as #1 and sections #1, #4, #7, and #10 are selected. Sections are post-fixed with 1% osmium tetroxide in water for 1 hour at room temperature (RT°), rinsed 3x5 min with water and contrasted "en bloc" for 1 hour at RT° with 2% aqueous uranyl acetate. After rinsing, sections are progressively dehydrated with 50%, 70%, 90%, and 100% ethanol solutions for 2x5 min each. Final dehydration is achieved by immersing the sections twice in 100% acetone for 10 min. Embedding is performed in epoxy resin (Embed 812, EMS, Euromedex, France) overnight in 50% resin / 50% acetone at 4°C followed by 2x2 h in pure resin at RT°, and polymerization is achieved at 56°C for 48 h in a dry oven. Semi-thin sections (0.5 μm-thick) are collected with an ultramicrotome UC7 (Leica, Leica Microsystèmes SAS, France) and stained with 1% toluidine blue in 1% borax buffer. Ultra-thin sections (70 nm-thick) are contrasted with Reynold's lead citrate (Reynold ES, 1963), and observed with a transmission electron microscope (HITACHI 120 kV HT 7700), operating at 70 kV. Images (2048x2048 pixels) are acquired with an AMT41B camera (pixel size: 7.4 μm x 7.4 μm).

## *2.4. Machine learning*

### *2.4.1 Random Forest*

Due to their robustness to noise as well as easiness of tuning (Criminisi et al., 2011), random forests are widely used as regression or classification techniques in the medical field (Alexander et al., 2017, Geremia et al., 2011, Nedjati-Gilani et al., 2017). A random forest is an ensemble technique, built of a collection of decision trees, called *weak learners*. A random forest regressor makes predictions by averaging the answers of all its decision trees, which are individually trained through a technique called *bagging*. This technique ensures the diversity of the trees by training each tree on a different random training subset. The randomness and diversity of the trees ensure their robustness to noise and good generalisation, resulting in the RF acting as a *strong learner* (Breiman, 2001). Here, we build an RF regressor that learns a mapping between the synthetic training database of DW-MRI signals/features and the ground truth microstructure parameters of the corresponding substrates. The mapping is learnt through a greedy splitting process of the input space (the synthetic signals/features) guided by the associated tissue parameters provided as labels during training.

There are two important parameters that need to be optimised to improve the learning performance of an RF: the number of trees and the maximum tree depth. The number of trees determines the smoothness of the decision boundary, and the tree depth parameter specifies the maximum levels that each decision tree can have. Too large a value can lead to overfitting while too low a value leads to underfitting, depending on the complexity of the data. Here, we run preliminary experiments and optimise these two parameters for our task in order to maximise the performance of our model.

### 2.4.2 Training and testing

We implement an RF regressor using the scikit-learn open source Python toolkit (Pedregosa et al., 2011). Following preliminary experiments, we build an RF with 200 trees of maximum depth 20 and bagging, as the setting the maximises the performance of the model. More general implementation details can be found at *http://scikit-learn.org/*. We train the RF for a multi-parameter regression task: we estimate the intra-axonal exchange time $\tau_i$ together with the intra-axonal volume fraction *f* and the intrinsic diffusivity *d*. Unlike the approach in Nedjati et al. (2017), we do not fit the axon radius index (Alexander et al., 2010) due to the lack of sensitivity of the signal to this parameter for our imaging protocol (Drobnjak et al., 2016).

The dimensionality of our synthetic databases is 11,000 by 345 for the signal database and 11,000 by 375 for the feature database. We set the size of training set to 11,000 as we did not find any improvements in performance above this number. The length of each synthetic training sample is reduced further during training according to the number of b shells selected in each training scenario. We train and test the RF on the synthetic databases using the associated ground truth parameters as labels for the supervised regression task. When predicting the parameter maps for the in-vivo data, we train the RF using the noisy databases as they are a more accurate representation of the in-vivo data. We split our synthetic database into a training set of 9,500 randomly selected signal/feature vectors and a test set formed of the remaining previously unseen 1,500 signal/feature vectors. As shown in Nedjati et al. (2017), the RF is not biased by the random selection of the training data as long as there is sufficient coverage of the parameter range, which we also ensure.

We adopt two different training scenarios for our experiments. The first one uses the raw signals database for training, and the second one uses the rotationally invariant features database obtained as in *Section 2.2.2*. While the first approach builds a direct mapping between the raw signals and the ground truth microstructure parameters, the second approach introduces an additional step of model fitting and constructs a mapping between DT and SH features of the raw signals and the microstructure parameters of interest.

## *2.5 Experiments*

### *2.5.1 Sensitivity analysis*

To ensure there is enough information in the data, we investigate the sensitivity of our PGSE protocol to the intra-axonal exchange time by looking at the range of $\tau_i$ values for which the DW-MRI signal can be distinguished from that of an impermeable substrate. For this, we consider two synthetic substrates representative of mouse white matter tissue, with the following properties: the mean axonal diameter $\mu_D$=0.4$\mu m$ and $\mu_D$=2$\mu m$, mimicking small and large axons in the CC (Barazany et al., 2009), the intra-axonal volume fraction $f$=0.7 (Barazany et al., 2009), and the intrinsic diffusivity=1.2$\mu m^2 ms^{-1}$ (Wu et al., 2008). These substrates are a good representation of our in-vivo mice data, as shown by the histological measurements of $\mu_D$ in *Section 3.5*, all within the range of the gamma-distributions above. Using the Camino toolbox, we generate synthetic signals for each substrate and different values of δ, Δ and G, corresponding to the b shells in our PGSE protocol. The diffusion gradients are set perpendicular to the cylinders in the substrate to maximise sensitivity to $\tau_i$. We investigate whether exchange time effects can be detected in the signal by looking at the difference in the normalised DW-MRI signal between impermeable ($\tau_i$=∞) and permeable substrates. Moreover, we analyse the effect of noise by looking at a range of different SNRs: SNR=∞, SNR=40 and SNR=20, where SNR=40 corresponds to the level of noise present in our in-vivo data. By using synthetic substrates representative of our in-vivo data and the same imaging protocol, we expect the analysis in this section to provide an indicative range of exchange time values for which there is reasonable sensitivity in our in-vivo data.

### *2.5.2 Shell selection*

As our imaging protocol uses an explorative range of imaging parameters, we select the b shells that maximise the performance of our RF model with respect to $\tau_i$. For this, we evaluate the performance of our RF model for every possible combination of 4, 9 and 16 shells out of the 25 in our protocol. We first evaluate combinations of 4 shells using as a benchmark the 4-shell STEAM protocol (Nedjati et al., 2017) optimised (Alexander, 2008) for a two-compartment model with exchange and biophysically plausible tissue parameters. As there are 12,650 possible combinations of 4 shells, we train the RF 12,650 times, once on each different shell combination. Then, for each training scenario corresponding to a unique combination of shells, we compute the correlation coefficient $R^2$ for $f$, $\tau_i$ and $d$ between the ground truth and the predicted values in the test set. Finally, we sort the different shell combinations according to their $R^2$ score for $\tau_i$ and choose the combination with the highest score as the one that maximises the performance of the model.

Furthermore, we investigate the effect of increasing the number of shells used for training. For this, we also look at combinations of 9 shells, as the minimum number of shells required to sample independently every unique G and Δ value in our PGSE protocol. Additionally, we look at combinations of 16 shells as a middle value between the 9-shell and the full protocol scenario. For this analysis, we use the synthetic feature-based dataset described in *Section 2.2*. Finally, we investigate the effect of noise on the performance of our model. For this, we look at a range of different SNRs:

SNR=∞, SNR=40 and SNR=20, where SNR=40 corresponds to the level of noise present in our in-vivo data.

### 2.5.3 Synthetic experiments

To assess the quality of the RF estimates after training is completed, we compute the Pearson correlation coefficient $R^2$ between the ground truth values and the RF predictions of the parameters in the previously unseen test set. To evaluate any potential bias in the estimates, we use Bland-Altman plots showing the mean of the predicted and ground truth values against their difference. We first analyse the performance of the model on the noise-free synthetic databases to establish a benchmark given our data and imaging protocol. Next, we apply our machine learning model to the SNR=40 database for a more accurate approximation of the performance we expect, given the noise present in our in-vivo data. For each experiment, we analyse both training scenarios outlined in *Section 2.4.2* (signal-based and feature-based) to test whether there are any differences in performance between the two approaches.

### 2.5.4 In-vivo imaging experiments

Before generating in-vivo parameter maps using our trained machine learning model, we first perform a data quality match to check that the in-vivo data is well represented by our synthetic training database. In addition to this, we investigate any potential bias in our in-vivo predictions of $\tau_i$ due to changes in the orientation dispersion by computing maps of the NODDI orientation dispersion index (ODI) (Zhang et al., 2012) using the NODDI Matlab (The MathWorks, Inc, Natick, MA) Toolbox[2]. Using the Camino toolbox, we additionally generate DTI maps at b=1500$s/mm^2$ of axial diffusivity (AD), fractional anisotropy (FA) and radial diffusivity (RD) as measures of tissue properties that can be compared with already published works in cuprizone model (Boretius et al., 2012, Song et al., 2005, Zhang et al., 2012b).

Using the random forest trained on the noisy database, we generate parameter maps for the CCs of the 16 mice for three parameters of interest: $\tau_i$, $f$ and $d$. To investigate the difference between the two groups (CPZ and WT), we compute box-and-whisker plots of region-specific comparisons between WT (8 mice) and CPZ (8 mice) for the DTI and NODDI metrics as well as for the RF predictions. Statistical significance is assessed by a two-tailed t-test, considering p-values<0.05. We run these experiments using the signals database. The Camino feature extraction of the in-vivo data did not produce histologically plausible results for the shells with very high gradient strengths (G>300mT/m) in our protocol, and we therefore exclude this approach from the analysis in this section. We discuss the potential explanations and the implications of this in *Section 4.1*.

### 2.5.5. Correlation with post-mortem analysis

From the electron microscopy (EM) samples obtained as described *Section 2.3.2*, we estimate the mean and standard deviation of the $g_{ratio}$, myelin thickness, axonal diameter and the intra-axonal volume fraction of the WT and CPZ mice. The stereological analysis is performed in isolated regions of the CC (genu, body and splenium), where 4 random sections with uniform distance are quantified per animal (*Figure 2B*), with 30 randomly located images per region and per animal acquired at 62000X magnification. For volume fraction (*VF*) we proceed according to the Delesse principle (Mouton, 2002): volume fractions are calculated by dividing the total number of points hitting the structure (*P(Y)*) by the total number of points hitting the reference volume (*P(ref)*), following the equation:

$$VF(Y, ref) = \frac{\sum_{i=1}^{m} P(Y)_i}{\sum_{i=1}^{m} P(ref)_i} \qquad (1)$$

---
[2] http://mig.cs.ucl.ac.uk/index.php?n=Tutorial.NODDImatlab

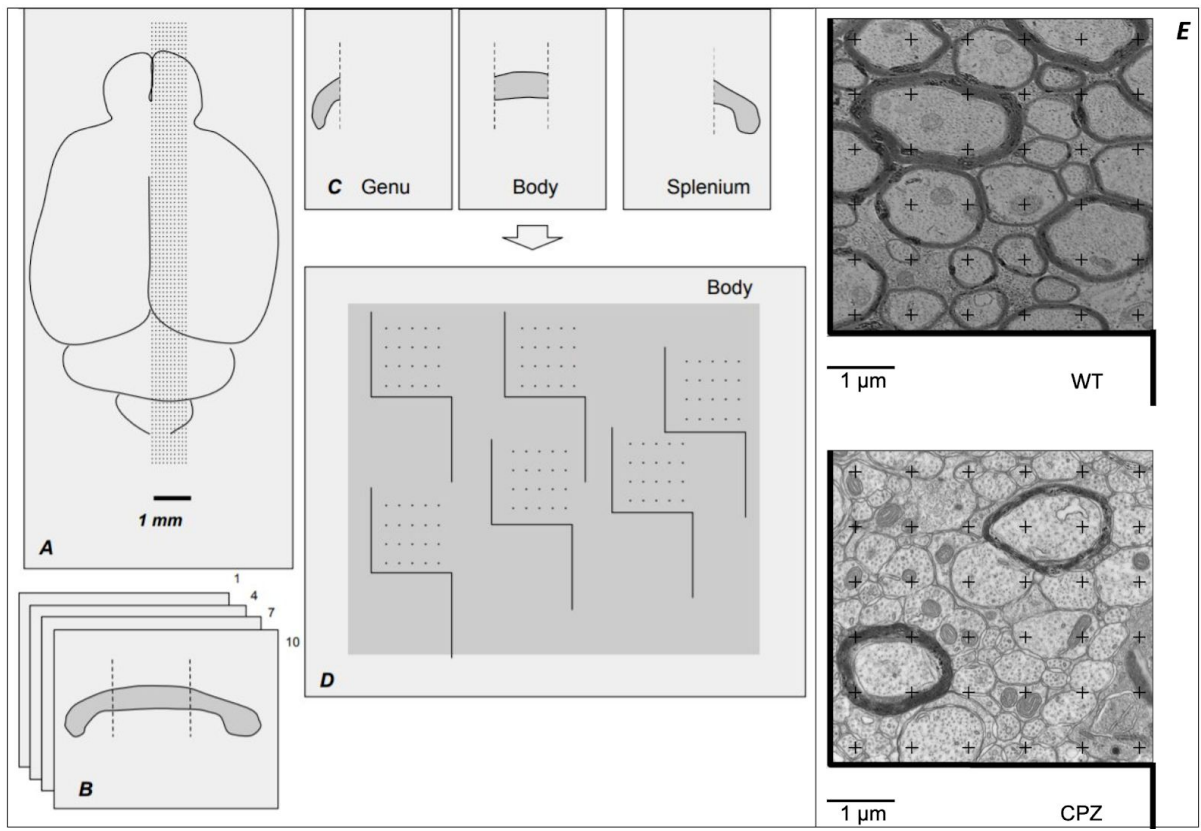

*Figure 2:* Schematic pipeline of the stereological analysis to compute $g_{ratios}$ and axonal diameters in the corpus callosum of the mice. First, ten sagittal sections are cut from medial to lateral lines (**A**) and four sections are selected for analysis (**B**). The genu, body and splenium are then extracted (**C**), and for each of the three ROIs 30 random regions are selected for stereological analysis using point grids (**D**). The point grids are then overlaid onto a tissue sample as illustrated in figure **E**, where each of the 36 crosses represents an area of 0.5μm². The point grids in **E** are used for quantification of the WT and CPZ mice.

A grid of 36 regularly spaced crosses (*Figure 2E*) is generated with Fiji, an open-source platform for biological image analysis (Schindelin et al., 2012). Each of the 36 crosses represents an area of 0.5 μm². Axonal fibers are judged to be in transverse section by the orientation of axonal microtubules. Stereological analysis provides Myelin Volume Fractions (MVF), Axon Volume Fractions (AVF), and the total Axon Volume Fractions (tAVF), which includes both myelinated and unmyelinated axons. Total Axon Count (tAxCount) is manually quantified. The $g_{ratio}$ of myelinated fibers is then calculated following equation Eq. 2:

$$g_{ratio} = \sqrt{\frac{AVF}{(MVF+AVF)}} \qquad (2)$$

The mean axon diameters (DAX) are calculated following Eq. 3:

$$DAX = 2\times\sqrt{\frac{(tAVF\times surface)}{(\pi\times tAxCount)}} \qquad (3)$$

The surface is the area of each cross (0.5 µm²). The outliers induced by the non-perpendicular axons in the images are not taken into consideration. From the $g_{ratio}$ and the DAX, myelin thickness is computed as:

$$\text{myelin thickness} = \frac{DAX}{2g_{ratio}}(1 - g_{ratio}) \qquad (4)$$

We compare the predictions of the RF with the EM measurements by computing the group-wise mean in the CC ROIs of the myelin thickness and intra-axonal volume fraction (VF) and looking at the correlation between these and the RF estimations for $\tau i$ and $f$.

## 3. Results

### 3.1 Sensitivity analysis

*Figure 3* shows the range of exchange time values for which the DW-MRI signal $S(\tau_i)$ can be distinguished from that of an impermeable substrate $S(\tau_i=\infty)$ in the presence of noise. For this, we calculate the change in signal $|S(\tau_i=\infty)-S(\tau_i)|$ between a permeable $S(\tau_i)$ and an equivalent impermeable $S(\tau_i=\infty)$ substrate. To illustrate practically achievable sensitivities, we plot this difference against three noise levels, denoted by the black plane: SNR=∞ (1st column), SNR=40 (2nd column) and SNR=20 (3rd column). *Figure 3A* illustrates the results for a substrate mimicking large axons in the white matter ($\mu_D$=2µm), while *Figure 3B* corresponds to a substrate with smaller axons ($\mu_D$=0.4µm). The second column shows that, for substrates with large axons (*row A*) and an SNR of 40, matching that of our in-vivo data, it is possible to distinguish exchange time effects for values of $\tau_i \leq$ 400ms. For substrates with small axons (*row B*), we can distinguish only permeable substrates with exchange times up to $\tau_i \leq$ 250ms. As expected, when the SNR drops to 20, it becomes harder to distinguish between impermeable and permeable substrates. This trend can be observed in the 3rd column, where the range for distinguishable permeable substrates narrows from $\tau_i \epsilon$ [0, 400]ms to $\tau_i \epsilon$ [0, 200]ms for large axons and from $\tau_i \epsilon$ [0, 250]ms to $\tau_i \epsilon$ [0, 140]ms for small axons.

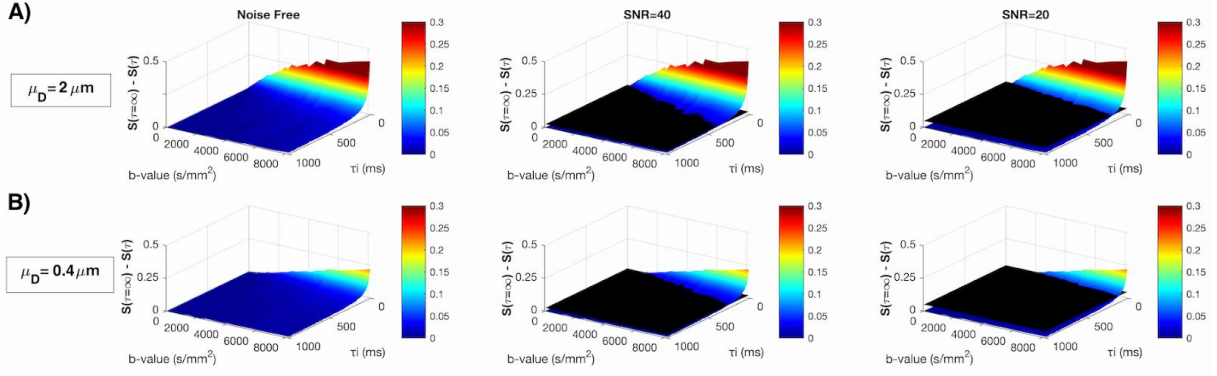

*Figure 3:* Exchange time ranges over which impermeable and permeable substrates can be distinguished for different noise levels and tissue substrates. Figure shows the difference in the DW-MR normalized signal between impermeable ($\tau_i=\infty$) and the equivalent permeable ($\tau_i \in [2, 1000]$ms) substrates. *Figure 3A)* shows results for a substrate with intra-axonal volume fraction $f = 0.7$ and mean axonal diameter $\mu_D=2\mu m$, representing large axons in the mouse brain. *Figure 3B)* illustrates the results for a substrate with $f = 0.7$ and $\mu_D=0.4\mu m$, mimicking small axons in the brain. The level of signal detectability is displayed for three SNR levels ($\infty$, 40 and 20), represented by the black planes, below which any change in signal is undetectable.

## 3.2 Shell selection

As our original 25-shell PGSE protocol uses an explorative range of imaging parameters, we optimise it with respect to the exchange time by choosing the shells most sensitive to this parameter (see *Section 2.5.2* for further details). In *Figure 5,* each point on the *x-axis* represents one unique shell combination and the corresponding *y-axis* value indicates the $R^2$ score when the RF is trained on that particular shell combination. For example, the *x-axis* in *Figure 5A* will have 12,650 points, each one corresponding to one of the 12,650 unique 4-shell combinations. As we are interested in the performance of the model with respect to $\tau_i$ (1$^{st}$ column), we rearrange the shell combinations in increasing order according to their $R^2$ for $\tau_i$. This results in a monotonically increasing curve for $\tau_i$, as seen in the first column. For $f$ (2$^{nd}$ column) and $d$ (3$^{rd}$ column), we keep the x-axis ordering consistent with the results for $\tau_i$ in the 1$^{st}$ column.

The $R^2$ scores curves in the 1$^{st}$ column of *Figure 5* show that only a limited number of shell combinations have a good correlation coefficient and are optimal for estimating $\tau_i$, while the $R^2$ scores in the 2$^{nd}$ and 3$^{rd}$ column show that the majority of shell combinations provide good estimates of $f$ and $d$. For example, in the noise free (blue curves) 4-shell case in the top row, we notice that the difference in $R^2$ score for $\tau_i$ between the best and the worst performing shell combinations is approximately 0.5. In contrast, this difference is much narrower for $f$ and $d$: ≈0.2 for $f$ and ≈0.1 for $d$. We observe the same trends for SNR=40 (orange) and SNR=20 (green), however, the differences in $R^2$ scores between the worst and the best performing shell combinations are smaller due to the presence of noise.

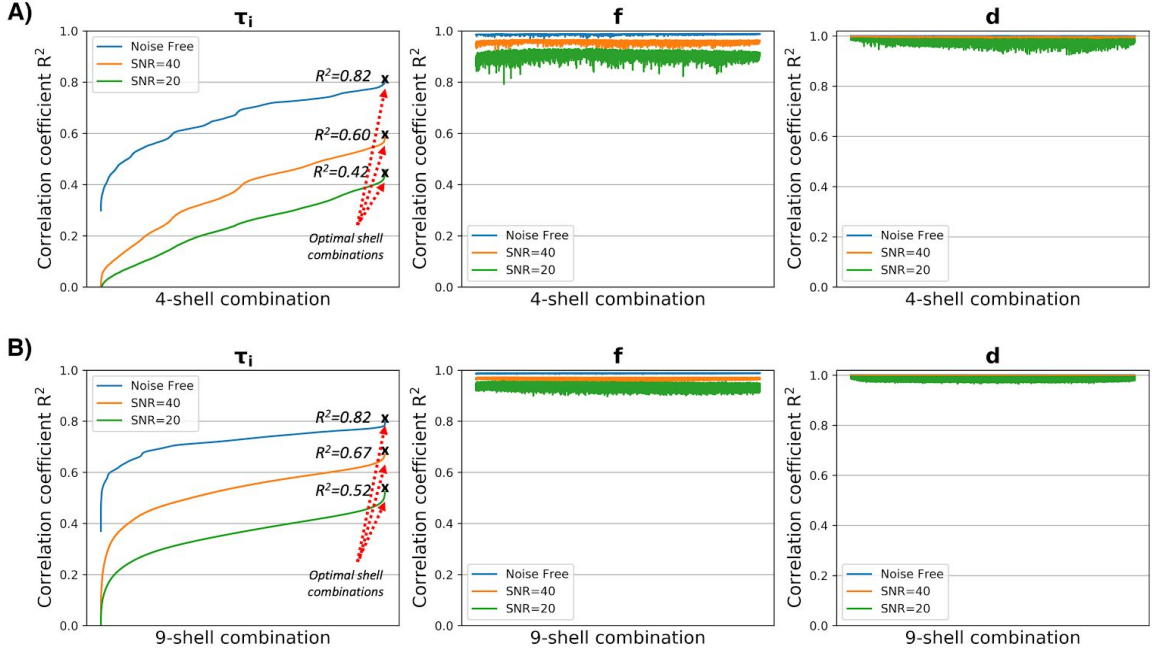

***Figure 5:*** *Performance of the RF model trained on different combinations of 4 **(A)** and 9 **(B)** shells. In the first column, each curve shows the $R^2$ score (y-axis) of the RF trained on a different combination of shells (x-axis). The shell combinations are sorted in increasing order according to their $R^2$ score. The second and third columns show the $R^2$ scores for f and d, which are sorted according to the order of the first column. We show the results for three levels of noise: SNR=∞ (blue curve), SNR=40 (orange curve) and SNR=20 (green curve). The $R^2$ score for $\tau_i$ is calculated only for values ≤400ms as this is the range over which we are sensitive to this parameter (see Section 3.1).*

By comparing the best $R^2$ scores on the blue curves in *Figure 5A* and *Figure 5B*, we can see that there is no difference in performance in the noise free scenario between using the best combination of 4 or 9 shells. However, this changes with the addition of noise. For example, for SNR=40 (orange curves) the $R^2$ score of the best 9-shell combination is 0.67, 0.07 higher than for the best 4 shells. This trend is similar for SNR=20 (green curves), with a difference of 0.1 between 9 and 4 shells. For the 16-shell scenario, we found no improvement in performance over using 9 shells.

*Figure 5* also shows the effect of noise on the estimation of each parameter. As expected, the addition of noise results in lower $R^2$ scores, a trend that holds for all parameters and across the 4 and 9-shell case. However, the estimation of $\tau_i$ is the most affected by the presence of noise: the maximum correlation coefficient drops from 0.82 in the noise free case to 0.67 for SNR=40 and even further to 0.52 for SNR=20. For $f$ ($2^{nd}$ column), the effect of noise is considerably smaller: $R^2$ drops from 0.99 for SNR=∞ to 0.94 for SNR=20. The estimation of the intrinsic diffusivity $d$ is very robust to noise: the correlation coefficients remaining very high (0.99) even when training the model on the SNR=20 dataset. Furthermore, we found that all the top 100 combinations contain the two highest b-values shells (6,803 and 7,773 smm$^{-2}$) which contain the two longest Δs. Additionally, we find that high b-value shells only maximise the performance of the random forest in combination with low b-value shells (775 and 930 smm$^{-2}$). For SNR=40 (orange curves), which we use when predicting on the in-vivo data, we found that the best combination of 9 shells sorted by b-value is [620, 775, 930, 1241, 1384, 2489, 4973, 6803, 7773]smm$^{-2}$ with an $R^2$ score of 0.67 and the best combination of 4 shells is [775, 930, 6803, 7773]smm$^{-2}$ with an $R^2$ score of 0.60. Since we are looking to optimise our framework for in-vivo estimation on the mouse data, we will run the in-vivo experiments using the best 9-shell combination in the SNR=40 scenario, as the noise level which matches our in-vivo data.

## 3.3 Synthetic experiments

*Figure 6* shows the RF results obtained using the feature (top row) and the signal (bottom row) noise free databases. To assess the quality of our fit, we display the results using Bland-Altman plots and colour each data point according to how close the estimates are to the predicted values. To aid visual interpretation, we cap the percentage error at ±50%. The mean difference between the ground truth and the predicted values is shown by the black line and the 95% upper and lower limits of agreement by the dashed lines. For all three parameters of interest, we observe no overall estimation bias as the predictions are spread equally around the zero-difference black line. However, for $\tau_i$, the parameter recovery is not perfect and the Bland-Altman plots show an overestimation bias for small values of $\tau_i$ and an underestimation bias for large values. The $R^2$ scores show a strong correlation between the estimates of our model and the ground truth parameter values: $R^2_{\tau i}$=0.82/0.84 (features/signals database), $R^2_f$=0.99 (both databases), and $R^2_d$=0.99 (both databases). When assessing the model's performance with respect to the two training databases (features/signals), we observe no significant difference between the two approaches. The $R^2$ scores remain unchanged for *f* and *d* and show only a minor difference for $\tau_i$: $R^2_{features}$ = 0.82 / $R^2_{signals}$ = 0.84. The advantages of each approach are discussed further in *Section 4*. The noise-free results in *Figure 6* provide a benchmark performance of the model given our data and imaging protocol.

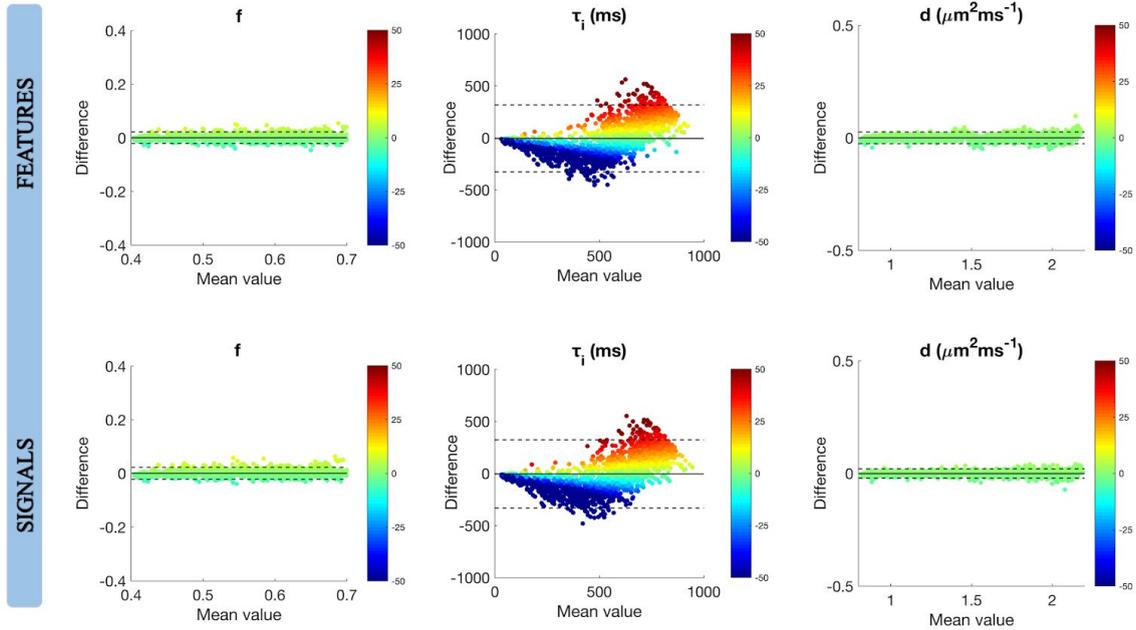

***Figure 6:*** *Bland-Altman plots for the RF predictions of f, $\tau_i$ and d using the features (top row) and signals (bottom row) noise-free simulated database. To aid visual interpretation, the plots are color-coded with the percentage error capped at ±50%.*

*Figure 7* shows the equivalent results for SNR=40. The presence of noise results in wider limits of agreement and affects differently the prediction of each parameter. The mean difference lines for all three parameters remain at zero, showing no general bias in the estimates. Intra-axonal volume fraction and diffusivity continue to be very well estimated and their correlation coefficients are only very mildly affected by the presence of noise: $R^2_f$=0.97 and $R^2_d$=0.99, equal for both training databases. In contrast to this, the presence of noise has a stronger effect on the estimation of $\tau_i$, resulting in a lower $R^2$ score and a more pronounced overestimation/underestimation bias for small and large values respectively. Despite this, we find that the RF works well within the sensitivity range computed in *Section 3.1*, with a very good correlation coefficient between the model's prediction and ground truth for $\tau_i \leq 400$ms ($R^2=0.68$). Outside this indicative sensitivity range, the correlation coefficient is very weak: $R^2= 0.07$ for $\tau_i \geq 400$ms. In line with the noise free case, we continue to see no significant difference between the signal and the feature approach: $R^2_{features} = 0.67$ / $R^2_{signals} = 0.68$.

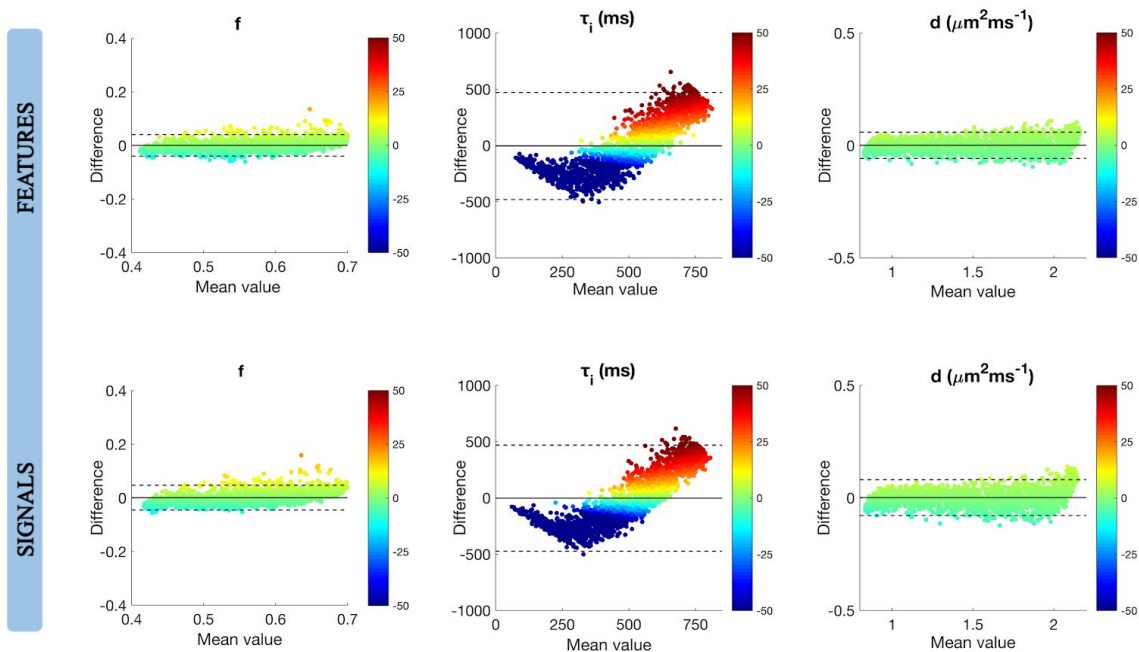

***Figure 7:*** *Bland-Altman plots for the RF predictions of f, $\tau_i$ and d using the features (top row) and signals (bottom row) simulated database with SNR=40, matching the noise level in our in-vivo data. To aid visual interpretation, the plots are color-coded with the percentage error capped at ±50%*

## 3.4 In-vivo imaging experiments

To show that our in vivo data is well represented by our synthetic training database, we perform a data quality match (*Figure 8*). We plot the signal intensity as a function of the angle between the diffusion gradients and the cylindrical fibres' axis θ (in degrees), for different diffusion gradient strengths ($G_{1-5}$=150-500 mT/m) and for Δ={10.8, 20.0}ms. We find a very good match between the simulated and in-vivo DW-MRI signals, demonstrating that our training data set is a good representation of the in-vivo mouse data.

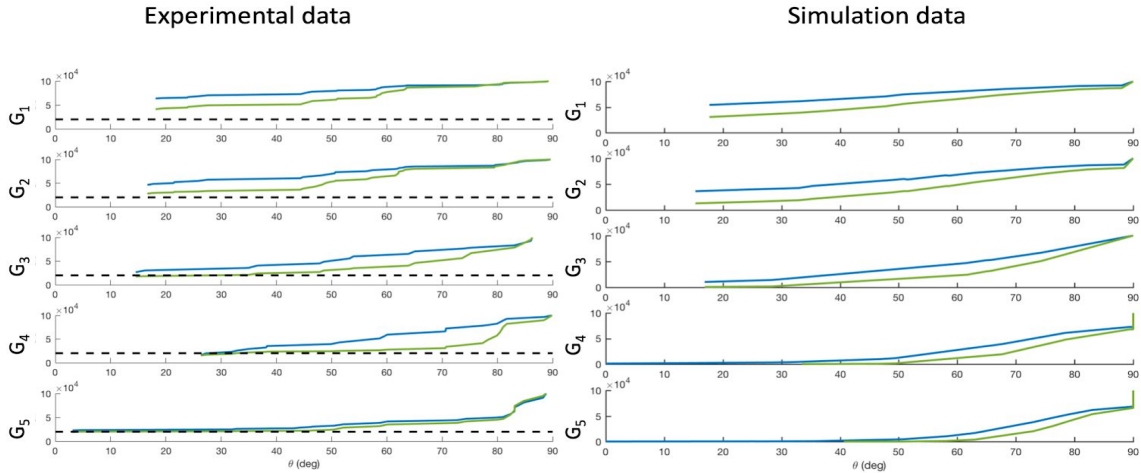

*Figure 8:* *Comparison between the in-vivo (left) and simulated (right) signal intensity as a function of the angle between diffusion gradients and the cylindrical fibres' axis θ (in degrees), for different diffusion gradient strengths ($G_{1-5}$=150-500mT/m) and two Δs: 10.8 ms (blue lines) and 20.0 ms (green lines). The dashed black line in the experimental data represents the noise level.*

*Figure 9* shows examples of DW-MRI b=0 images for a WT (*Fig. 9A*) and for a CPZ (*Fig. 9B*) mouse. We can observe the appearance of the CC in the CPZ scan is different from the WT, showing the effect of demyelination. *Figure 9C)* shows the three ROIs of the CC overlaid on the b=0 image of the WT scan. We manually define three ROIs on the CC masks of each mouse scan: splenium (S-CC), body (B-CC) and genu (G-CC). We then calculate the mean parameter estimates for NODDI (ODI), DTI (AD, RD, FA) and RF (f, $\tau_i$, d) in each ROI for every mouse, and study the differences between the WT and the CPZ groups. We present these results in the remaining of this section.

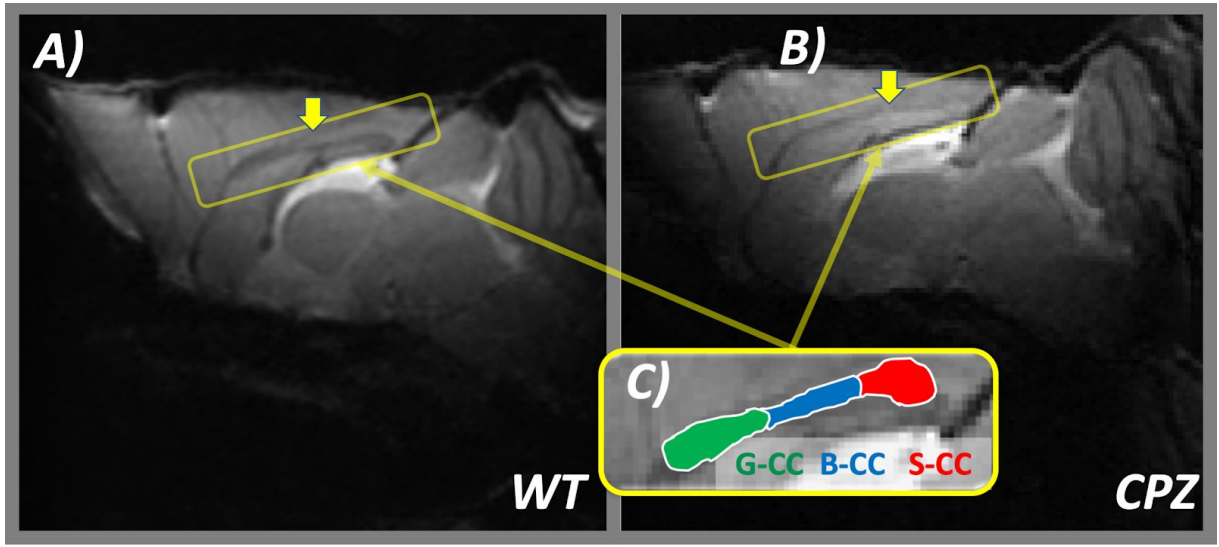

*Figure 9:* *Representative DW-MRI b=0 images of **A)** a WT mouse scan in our cohort and **B)** a CPZ mouse scan in our cohort. **C)** ROIs of the CC overlaid on the zoomed in b=0 image of the WT mouse scan. The three ROIs are genu (G-CC), body (B-CC) and splenium (S-CC). The yellow square indicates the region in which the CC is found.*

*Figure 10* shows CC maps for NODDI and DTI parameters for one exemplar healthy WT mouse (first column) and one exemplar CPZ mouse (second column). A visual inspection of the CC maps reveals no significant changes in ODI and AD, together with a significant increase in RD and decrease in FA.

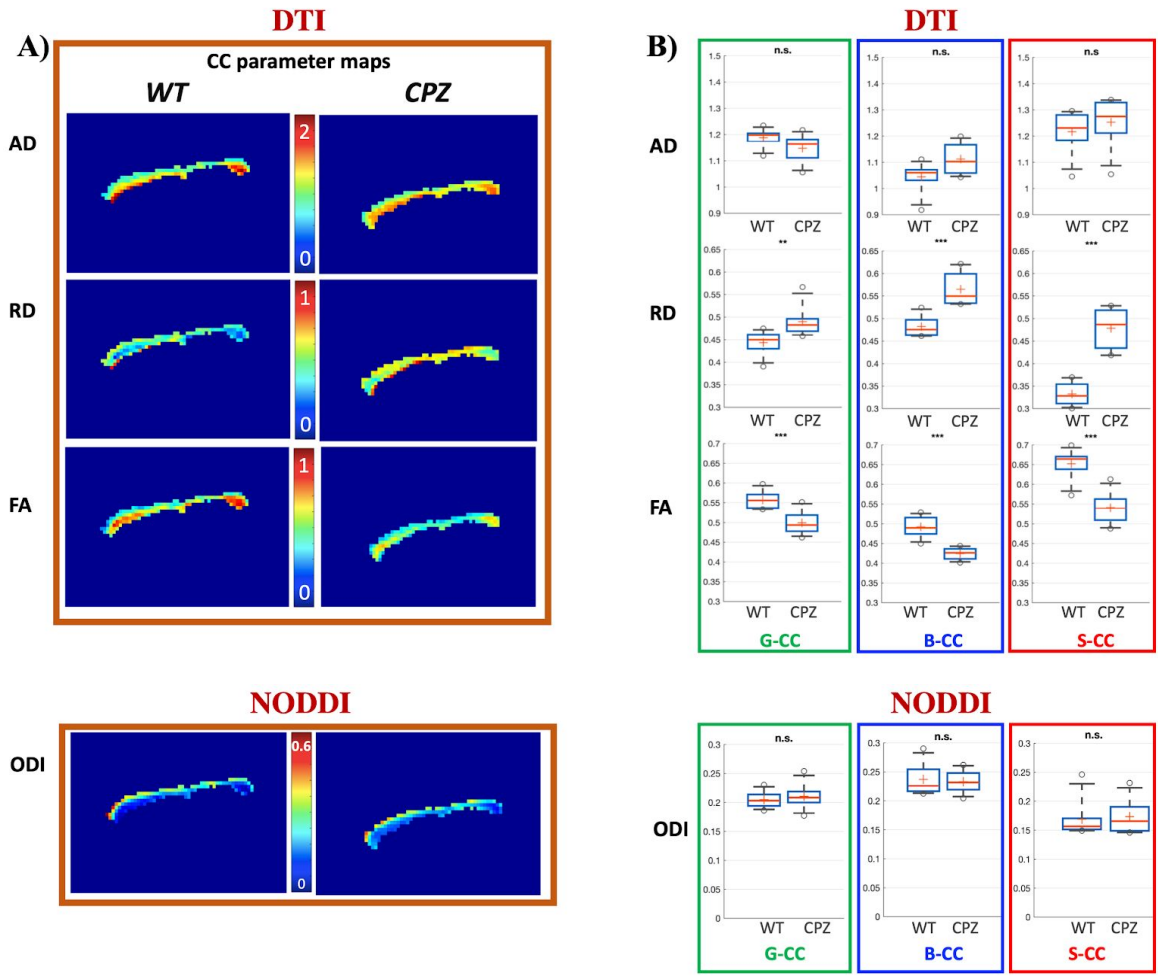

*Figure 10:* **A)** *Parametric maps of the CC in a healthy WT mouse (first column) and a CPZ mouse (second column) obtained from conventional DTI at b=1500s/mm² and from NODDI ODI.* **B)** *Box and whisker plots of region-specific comparison between WT ($N_{mice}$=8) and CPZ ($N_{mice}$=8). DTI metrics (AD, RD, FA) are evaluated within the genu (G-CC), body (B-CC) and splenium (S-CC) of the CC. Statistical significance is assessed by using a 2-tailed t-test with equal variance and significance level: \*=0.01, \*\*=0.005, \*\*\*=0.001. 'n.s.' stands for non-significant.*

We observe the same trends in the DTI and NODDI parameters at group level, as shown in *Figure 10B)*. We illustrate the difference between the WT group ($N_{mice}$=8) and the CPZ group ($N_{mice}$=8) through box and whisker plots in the three ROIs of the CC: genu (G-CC), body (B-CC) and splenium (S-CC). We find no statistically significant difference in NODDI ODI between the two groups in the three regions of the CC, ruling out a potential bias in our predictions of $\tau_i$ due to changes in the orientation dispersion. The DTI estimates show negligible changes in AD, a significant increase in RD and a significant decrease in FA.

The in-vivo RF predictions of $f$, $\tau_i$ and $d$ obtained using the raw signal database are presented in *Figure 11*.

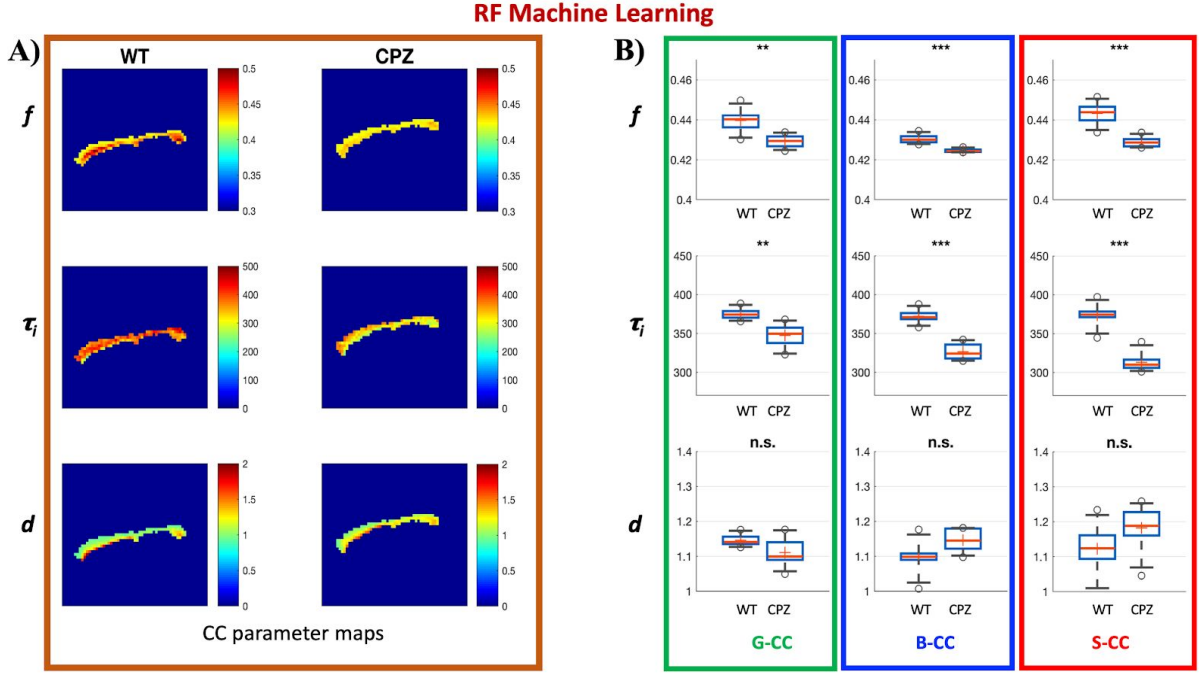

***Figure 11: A)*** *Parametric maps with the RF predictions for f, $\tau_i$ and d in the CC of a healthy WT mouse (first column) and a CPZ mouse (second column).* ***B)*** *Box and whisker plots of region-specific comparison between WT ($N_{mice}$=8) and CPZ ($N_{mice}$=8). RF predictions for f, $\tau_i$ and d are computed independently for all voxels within the genu (G-CC), body (B-CC) and splenium (S-CC) of the CC. Statistical significance was assessed by using a 2-tailed t-test with equal variance and significance level: \*=0.01, \*\*=0.005, \*\*\*=0.001. 'n.s.' stands for non-significant.*

The parametric CC maps shown in *Figure 11A)* correspond to the same WT mouse (first column) and CPZ mouse (second column) in *Figure 10A)*. The CC maps show a statistically significant decrease in *f* (first row) and $\tau_i$ (second row), and no significant change in *d* (third row). To provide a more quantitative analysis, we plot the box and whisker plots of region-specific parameter comparisons between the WT and the CPZ group over the three CC ROIs (*Figure 11B)*. The trends observed visually in *Figure 11A)* hold for the group-wise quantitative comparison (WT versus CPZ): we observe statistically significant decreases in *f* and $\tau_i$ and a negligible increase in *d*. These trends are consistent across all three regions of the CC. The mean and standard deviations of the RF parameter predictions for each ROI are reported in *Table 2*.

|  | *f* |  | $\tau_i$ |  | *d* |  |
|---|---|---|---|---|---|---|
|  | ***WT*** | ***CPZ*** | ***WT*** | ***CPZ*** | ***WT*** | ***CPZ*** |
| **S-CC** | 0.443 (0.005) | 0.428(0.003)*** | 370 (0.01) | 310 (0.01) *** | 1.12 (0.07) | 1.18 (0.07) |
| **B-CC** | 0.430 (0.002) | 0.424(0.001)*** | 370 (0.01) | 330 (0.01) *** | 1.10 (0.05) | 1.15 (0.03) |
| **G-CC** | 0.440 (0.006) | 0.429(0.003)** | 380 (0.01) | 350 (0.02) ** | 1.15 (0.02) | 1.11 (0.04) |

***Table 2:*** *Mean and standard deviation of RF predictions for f, $\tau_i$ and d in the three CC ROIs for the WT and CPZ group. CPZ regions that are statistically different from WT regions are marked with \* for p<0.01, \*\* for p<0.005 and \*\*\* for p<0.001.*

## 3.5 Correlation with post-mortem analysis

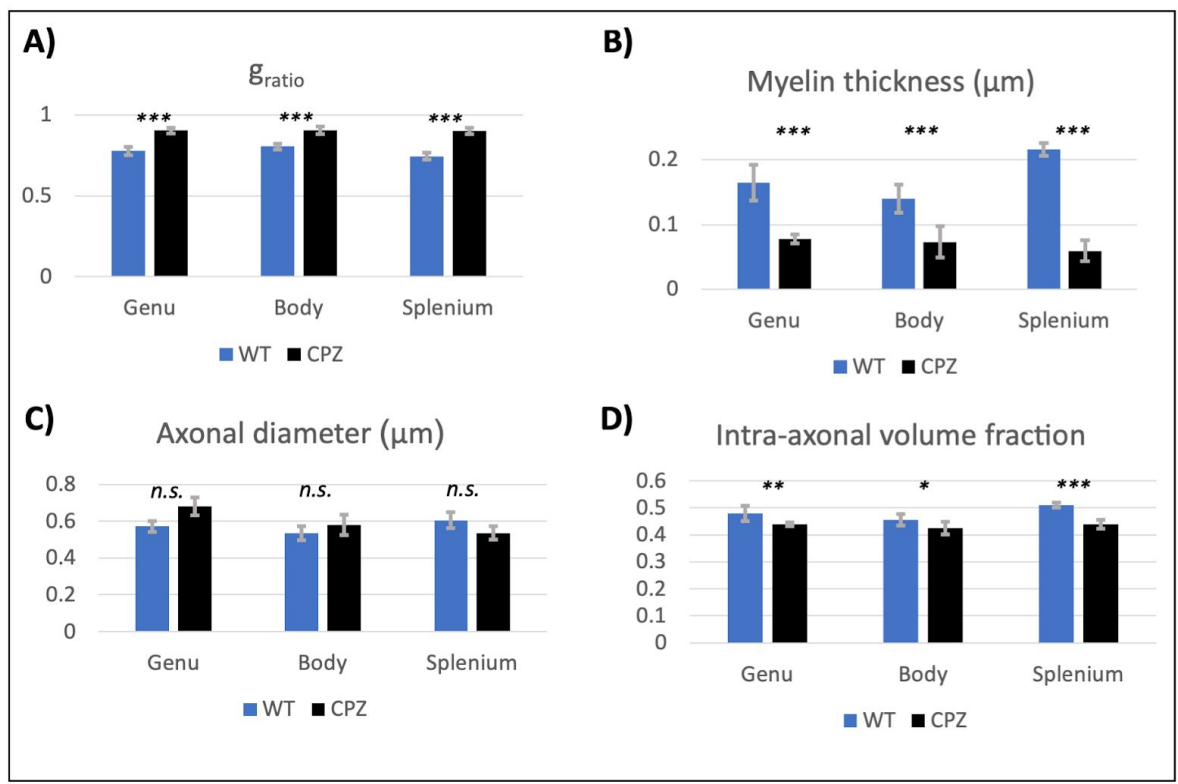

*Figure 12: Histology results. The mean and the standard deviation of the EM measurements in the splenium, body and genu of the CC for the cohort of WT (blue) and CPZ (orange) mice: the $g_{ratio}$ (A), myelin thickness (B), mean axonal diameter (C) and intra-axonal volume fraction (D).*

The histological EM measurements in the splenium, body and genu of the CC over the cohort of WT (blue) and CPZ (black) mice are reported in the histograms of *Figure 12*. Our histological data shows no axonal size changes (*Figure 12C*) and no significant axonal loss (data not shown here) between the two cohorts. The axonal diameter measurements in *Figure 12C* do not take into account the commonly accepted shrinkage factor of 30% (Barazany et al., 2009, Innocenti et al., 2015), after which the differences between the two groups continue to remain statistically non-significant. We also find a statistically significant decrease in myelin thickness (*Figure 12B*) correlated with an increase in the $g_{ratio}$ (*Figure 12A*) and a decrease in the intra-axonal volume fraction (*Figure 12D*).

Next, we study the correlation between these changes and the predictions of the RF model in *Figure 13*. We assess the statistical significance of the linear correlation between $\tau_i$ and myelin thickness from EM with a two-tailed t-test by looking at the mean and the standard deviation of each CC ROI of the WT (blue squares) and CPZ (black circles) group (*Figure 13A*). We find a Pearson linear correlation coefficient $\varrho$ of 0.82 and a p-value < 0.05 for $\tau_i$, showing a good correlation between the RF estimates of the exchange time from DW-MRI *(y-axis)* and histological measurements of myelin thickness *(x-axis)*.

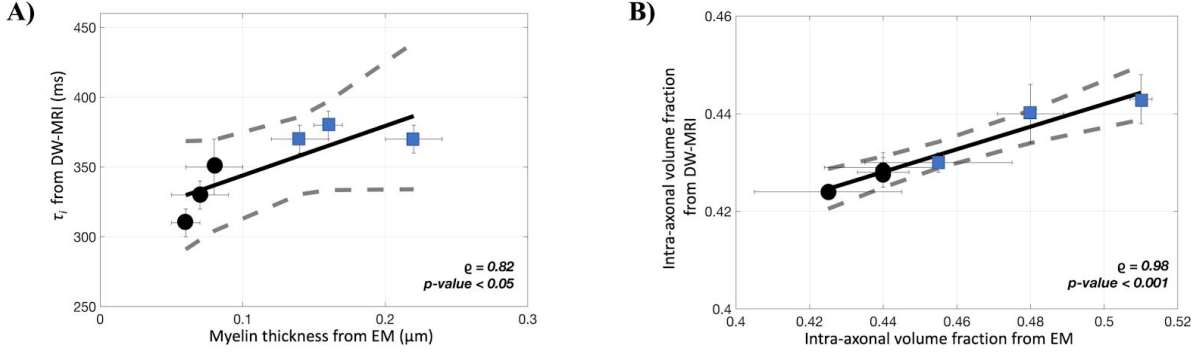

*Figure 13: Statistical significance and correlations between **A)** exchange time from DW-MRI (y-axis) and myelin thickness from EM (x-axis) and **B)** between intra-axonal volume fraction from DW-MRI (y-axis) and EM (x-axis). Each point represents the mean over one region of the CC for the WT (blue squares) and CPZ (black circles) group. Error bars indicate the standard deviation over the region.*

Similarly, we investigate the statistical significance of the linear correlation between intra-axonal volume fraction *f* as estimated from DW-MRI *(y-axis)* and EM measurements *(x-axis) (Figure 13B)*. We find a Pearson correlation coefficient $\varrho$ of 0.98 and a p-value<0.001, showing a strong correlation between the RF estimates and the histological measurements of the intra-axonal volume fraction.

## 4. Discussion

In this work, we focus on the experimental study of a random forest based computational model for axonal permeability estimation using an in-vivo cuprizone mouse model of demyelination. For this, we use Monte Carlo simulations of the DW-MRI signal and train our model to estimate microstructure parameters with a focus on the intra-axonal water exchange time $\tau_i$, a parameter inversely related to axonal permeability. Using synthetic substrates mimicking our in-vivo data, we show that our imaging protocol has good sensitivity to exchange times $\tau_i \leq 400ms$ for large axons (mean diameter of 2$\mu m$) and to $\tau_i \leq 250ms$ for small axons (mean diameter of 0.4$\mu m$) under the noise conditions of our in-vivo data (SNR=40). Following from this, we find that the RF model we developed works very well in this range: we find a good correlation between RF predictions and the ground truth for $\tau_i \leq 400ms$ ($R^2$=0.87 for SNR=inf and $R^2$=0.68 for SNR=40), and a weak correlation for $\tau_i > 400ms$ ($R^2$=0.3 for SNR=inf and $R^2$=0.07 for SNR=40). In our in-vivo imaging experiments, we find that the RF estimates of $\tau_i$ are within the sensitivity range and in line with literature values of the exchange time reported in healthy rat brain tissue (Prantner, 2008, Quirk et al., 2003). Furthermore, we find that the RF estimates of $\tau_i$ in the CPZ group are significantly lower than in the WT group, a finding that one would intuitively expect to see in a model of demyelination. Furthermore, we find that our intra-axonal volume fraction estimates in CPZ mouse are also significantly lower than in controls. These results are in strong agreement ($\varrho_f$ = 0.98 and $\varrho_{\tau i}$ = 0.82) with our EM histology results of myelin thickness and intra-axonal volume fraction. Finally, we show that potentially confounding factors such as axonal swelling and dispersion have a negligible effect. These results suggest for the first time, quantitatively and in-vivo, that machine learning based computational models could act as a suitable biomarker to detect and track changes in demyelinating pathologies.

## 4.1 Simulations

*Sensitivity analysis.* Our sensitivity analysis shows that our imaging protocol has good sensitivity for exchange times in the range $\tau_i \in [0, 400]$ms for substrates with large axons (mean diameter $\mu_D=2\mu m$) and in the range $\tau_i \in [0, 250]$ms for substrates with small axons (mean diameter $\mu_D=2\mu m$), under noise conditions matching that of our in-vivo data (SNR=40). Generally speaking, the noise in the data affects the sensitivity differently, depending on the mean axon diameter in the substrate. For substrates with large axons ($\mu_D=2\mu m$), the sensitivity halves from $\tau_i \in [0, 400]$ms for SNR=40 to $\tau_i \in [0, 200]$ms for SNR=20. For substrates with smaller axons ($\mu_D=0.4\mu m$), decreasing the SNR from 40 to 20 has a smaller effect on the sensitivity range, reducing it by 44% from $\tau_i \in [0, 250]$ms (SNR=40) to $\tau_i \in [0, 140]$ms (SNR=20). Furthermore, we found that the larger the axons in the substrate, the better the sensitivity range. Substrates with $\mu_D=2\mu m$ have a sensitivity range wider by 60% (for SNR=40) and by 43% (for SNR=20) than substrates with $\mu_D=0.4\mu m$.

*Shell selection.* To optimise the performance of the machine learning model, we explore the wide range of parameters in our PGSE protocol and select the best combination of 4 and 9 shells. We show that for our in-vivo data with SNR=40 the number of shells that maximises the performance of the model is 9, with the b-values [620, 775, 930, 1241, 1384, 2489, 4973, 6803, 7773]smm$^{-2}$ and an $R^2$ score of 0.67. When analysing the best combinations of 4 and 9 shells, we observe that they sample every value of $\Delta$ in our sequence, resulting in a combination of low and high b-values shells. This finding is in accordance with the optimised STEAM protocol in Nedjati et al. (2017), which contains two long $\Delta$ and two short $\Delta$ shells. This suggests that to maximise sensitivity to the intra-axonal exchange time, it is necessary to include a combination of short and long $\Delta$s.

We show that noise is an important factor for the performance of our model. We find that in the noise free case, it is sufficient to use only 4 shells as introducing more shells does not improve performance. However, in the presence of noise, we find that increasing the number of shells from 4 to 9 improves the R$^2$ score between the predicted and the estimated $\tau_i$. A potential explanation for this is that the addition of noise corrupts the information in each shell, and having more shells to corroborate information from helps the RF model learn better. Our analysis also reveals that increasing the number of shells above 9 does not offer any additional benefits even in the presence of noise. Moreover, we show that noise has a stronger effect on the estimation of $\tau_i$, for which the $R^2$ score drops from ≈0.85 in the noise free case to ≈0.5 for SNR=20. The estimation of *f* and *d* is considerably more robust: $R^2_{noise-free}=0.99$ versus $R^2_{SNR=20}=0.94$ for *f* and no drop for *d*. This suggests that SNR plays an important role in a protocol's suitability for permeability estimation using our approach.

*Feature extraction.* When extracting the rotationally invariant features from our synthetic signals, we obtain meaningful values for all shells in the synthetic data. When we apply the same method to in-vivo data, the feature extraction becomes difficult and does not give meaningful results for shells with higher gradient strength (above 300mT/m) and higher b-values. We believe that this difference is most likely due to the effect of fibre dispersion, present in the in-vivo data but not included in our simulations. As the gradient strength increases, the dispersed fibres would cause larger drops in the signal, as can also be seen in (*Figure 8*), where we notice that the drop in the signal intensity relative to the gradient direction is less prominent in the synthetic signals than in the in-vivo data.

*Synthetic data experiments.* The RF model predictions in the noise free case have very strong correlations with the ground truth values, providing an excellent benchmark performance for our model and imaging protocol (*f*: $R^2$=0.99, $\tau_i$: $R^2$=0.84 *d*: $R^2$=0.99). We show that the addition of noise with SNR=40, matching our in-vivo data, does not affect much the estimation of *f* and *d* (*f*: $R^2$=0.94, *d*: $R^2$=0.99), however, it has a stronger effect on the estimation of $\tau_i$. In line with our sensitivity results, for $\tau_i$<400ms the effect is present, however, the performance is still sufficiently good ($R^2$=0.68), while for $\tau_i$>400ms the performance of the model is severely affected ($R^2$=0.07).

In addition to this, we compare for the first time the signal and feature training approaches and show that there is no significant difference in the RF performance according to which database is used for training. This is a significant result as it shows that when extracting the rotationally invariant features from the raw signals we do not lose information that is essential for training our model. Consequently, we can use the features database without affecting the performance of our model. The advantage of a rotationally invariant feature approach is that it does not require the generation of a new library for every new acquisition protocol as long as the b-values and the TE of the protocols match. Nevertheless, as discussed above, caution should be applied with this approach when the acquisition protocol uses high gradient strengths (G≥300mT/m) and the SNR is low, such as conditions often found in pre-clinical setting, and then using signals database might be the preferable choice. On the other hand, in the clinical setting, imaging protocols have much lower gradient strengths and sufficient SNR to fit the DTI and SH model parameters in the feature extraction approach, and consequently, we expect the rotationally invariant feature approach to be a better choice (as used in Nedjati et al. (2017)). Irrespective of the training approach, we expect our model's performance to be similar in both the clinical and preclinical setting.

*4.2 In-vivo mouse data and correlation with post-mortem analysis*

Our data quality match shows that our synthetic training data is a good representation of the in-vivo data. Our DTI results show an increase in RD and a decrease in FA between the two groups. This could be explained by the breakdown of the myelin layer which allows water to diffuse more in the radial direction, leaving AD unchanged and having the overall effect of reducing FA. These changes in DTI metrics are in agreement with those reported in several literature studies of the CPZ mouse model of demyelination (Boretius et al., 2012, Song et al., 2005, Zhang et al., 2012b). Nevertheless, the DTI metrics provide only indirect measures of the underlying microstructural changes of the CPZ model.

Our RF predictions of $\tau_i$, on the other hand, provide a more direct and specific measure of permeability. We find that our estimations of $\tau_i$ in the healthy mice compare well with literature values. Studies on sphingomyelin membranes found in axonal membranes suggest values between 300 ms and 600 ms for axons with radii between 0.5 and 1 μm (Finkelstein, 1976). Contrast agent and relaxometry studies in the rat brain estimate the intracellular water exchange lifetime in the rat brain to be between 200ms (Prantner, 2008) and 550ms (Quirk et al., 2003). As accurate histology measurements of $\tau_i$ are not available due to tissue fixation altering the membrane permeability, we compare our estimates of $\tau_i$ with EM measurements of myelin thickness. We compute myelin thickness from myelinated axons only, and it includes both the effect of demyelination induced by CPZ and some remyelination that happens spontaneously in the CPZ model (Matsushima and Morell, 2001). We find a strong correlation between the RF predictions of $\tau_i$ and myelin thickness ($\varrho_{\tau i}$ = 0.82). This is in very good agreement with a recently published simulation work investigating the link between exchange time and myelin thickness (Brusini et al., 2019). Furthermore, our RF estimates of $d$ lie in the range 1-1.3$\mu m^2 s m^{-1}$, an expected range for the mouse CC (Wu et al., 2008) and our estimates for $f$ correlate very strongly with the EM intra-axonal volume fraction measurements ($\varrho_f$ = 0.98).

When comparing the two groups, we observe the following general trends: a statistically significant decrease in the intra-axonal volume fraction $f$ and in the intra-axonal exchange time $\tau_i$, together with a negligible increase in the intrinsic diffusivity $d$. We expect $f$ to be lower in the CPZ group as there is an increase in the extracellular space due to the breakdown of myelin. Demyelination is also thought to cause a decrease in the intra-axonal exchange time as the water molecules encounter less barriers when moving from the intracellular to the extracellular space. In line with this, the RF estimations of $\tau_i$ in the CPZ group are significantly lower than in the WT group.

The strong correlation between myelin thickness and the estimated $\tau_i$ suggests that demyelination could be one of the main factors behind our measured decrease in $\tau_i$. To strengthen this hypothesis, we analyse the potential confounding effect of other underlying processes. Our AD measurements from the DTI fit suggest that, if undulation or beading effects are present, they have a negligible effect (Budde and Frank, 2010, Nilsson et al., 2012, Palombo et al., 2018). We additionally rule out the effect of dispersion by measuring NODDI – ODI and showing that the differences between the two groups are not statistically significant. We also rule out the potential confounding effect of axonal swelling by looking at the statistically non-significant changes in axonal diameter as measured by EM. This, together with the measured changes in RD, FA and the RF estimations of $\tau_i$ suggest that demyelination is the main process underpinning our DW-MRI contrast. In particular, our histological data strongly supports $\tau_i$ as a biomarker directly related to the thickness of the myelin sheath, which suffers degeneration in demyelinating diseases such as Multiple Sclerosis.

### *4.3 Limitations*

*Tissue model.* One limitation of the present work is the simplicity of the white matter substrates used for the Monte Carlo simulations. Due to current limitations in our simulation system, we make several assumptions about the geometry of the tissue such as representing axons as non-abutting parallel cylinders. Future work should aim to train the machine learning model on more realistic simulations, which account for different effects such as myelin water (Harkins and Does, 2016, Brusini et al., 2019), axonal undulation (Nilsson et al., 2012), dispersion (Ginsburger et al., 2019, Ross Callaghan, 2019), neurons and glial cells (Palombo et al., 2018). Such effects, once included in the simulations, can easily be incorporated in the machine learning framework used in this paper. Using more complex and realistic simulations will also narrow down the gap between the synthetic and in-vivo data and increase the robustness of the parameter estimates. The gap between simulations and in-vivo data could also be addressed by using domain adaptation techniques, which can be integrated within other machine learning approaches such as neural networks.

*Sensitivity of the imaging protocol.* The sensitivity of our imaging protocol to the exchange time in the presence of noise, as we show in our simulation experiments, is not ideal. Although we performed some level of optimisation by choosing the most optimal shells in our large explorative data set, the sensitivity might have been better if we optimised the protocol using computational framework with respect to $\tau_i$ prior to imaging as done in Nedjati et al. Nevertheless, even with the protocol we use we can estimate values of $\tau_i \leq 400$ms, which is sufficient for the in-vivo mouse application we used in this paper. The machine learning model here can also easily be adapted to incorporate more specialised diffusion encoding sequences such as OGSE for more sensitivity to axon diameter (Drobnjak et al., 2016) or STEAM for longer diffusion times (Fieremans et al., 2016). An advantage of the machine learning framework in this paper is that it can easily incorporate changes in the Monte Carlo simulations and imaging protocol.

*RF model validation.* Another limitation that stems from our simulation system is the testing of the model on the same type of data as it is trained on. Nevertheless, it is worth mentioning that despite using the same tissue model, we test our model using previously unseen parameter values. Future work should include the training and testing of the machine learning model using different types of substrates once these become available.

## *4.4 Implications for clinical applications*

The extension of our approach to clinical systems can potentially be important for numerous white matter pathologies of the human nervous system. Here, we demonstrate the potential of our model in a cuprizone mouse model of demyelination, which is extensively used in the MS literature due to its close similarity to the demyelination and remyelination processes occurring in MS lesions (Ransohoff, 2012). This suggests that our approach can potentially be suitable for translation to the clinic, which has also been preliminary shown in Nedjati et al. (2017) on MS patients.

The clinical applicability of our approach extends to other myelin damaging pathologies such as spinal cord injury or leukodystrophies due to the hypothesised correlation between $\tau_i$ and the condition of the myelin sheath (Nilsson et al., 2013b, Ford and Hackney, 1997, Hwang et al., 2003). The current key limitation is the reduced sensitivity to the intra-axonal exchange time of clinically available imaging protocols. This can be addressed by using more specialised sequences such as the AXR sequence (Nilsson et al., 2013a) or optimised STEAM pulse sequences as in Nedjati et al. (2017), to which our framework can easily be adapted. With the continually increasing SNR in the clinical scanners, we expect the clinical applicability of this approach to also improve.

Our machine learning approach can be easily extended to a range of other intractable parameters such as undulation or extracellular space. Another important further development for clinical practice is the introduction of uncertainty measures on the predictions of $\tau_i$. Uncertainty could be included via a Bayesian approach as in Tanno et al. (2016) and it would help highlight areas of the brain where the predictions are less reliable due to unfamiliar signals. From a clinical perspective, these future developments can have a great impact on the understanding and diagnosis of neurological conditions of the white matter.

# Acknowledgements

This work was supported by EPSRC (EP/G007748, EP/I027084/01, EP/L022680/1, EP/M020533/1, EP/N018702-1), EPSRC-funded UCL Centre for Doctoral Training in Medical Imaging (EP/L016478/1), Spinal Research, UK MS Society, NIHR and NIHR UCLH Biomedical Research Centre (BRC).

The research leading to these results received funding from the programs 'Institut des neurosciences translationnelles ANR-10-IAIHU-06 and 'Infrastructure d'avenir en Biologie Santé' ANR-11-INBS-0006 and 'Translational research infrastructure for biotherapies in neuroscience' ANR-11-INBS-0011. We thank the ICM Quant Cellular Imaging platform for their assistance in EM.